\documentclass[a4paper,11pt]{article}

\usepackage{pos}
\usepackage{graphicx}
\usepackage{dcolumn}
\usepackage{bm}
\usepackage{amsmath}
\usepackage{float}
\usepackage{multirow}
\usepackage{slashed}
\usepackage{xcolor}
\usepackage{braket}
\usepackage{physics}
\usepackage{multirow}
\usepackage{gensymb}
\usepackage{mathtools,braket}
\usepackage{hyperref}

\def\orcid#1{\kern .08em\href{https://orcid.org/#1}{\includegraphics[keepaspectratio,width=0.7em]{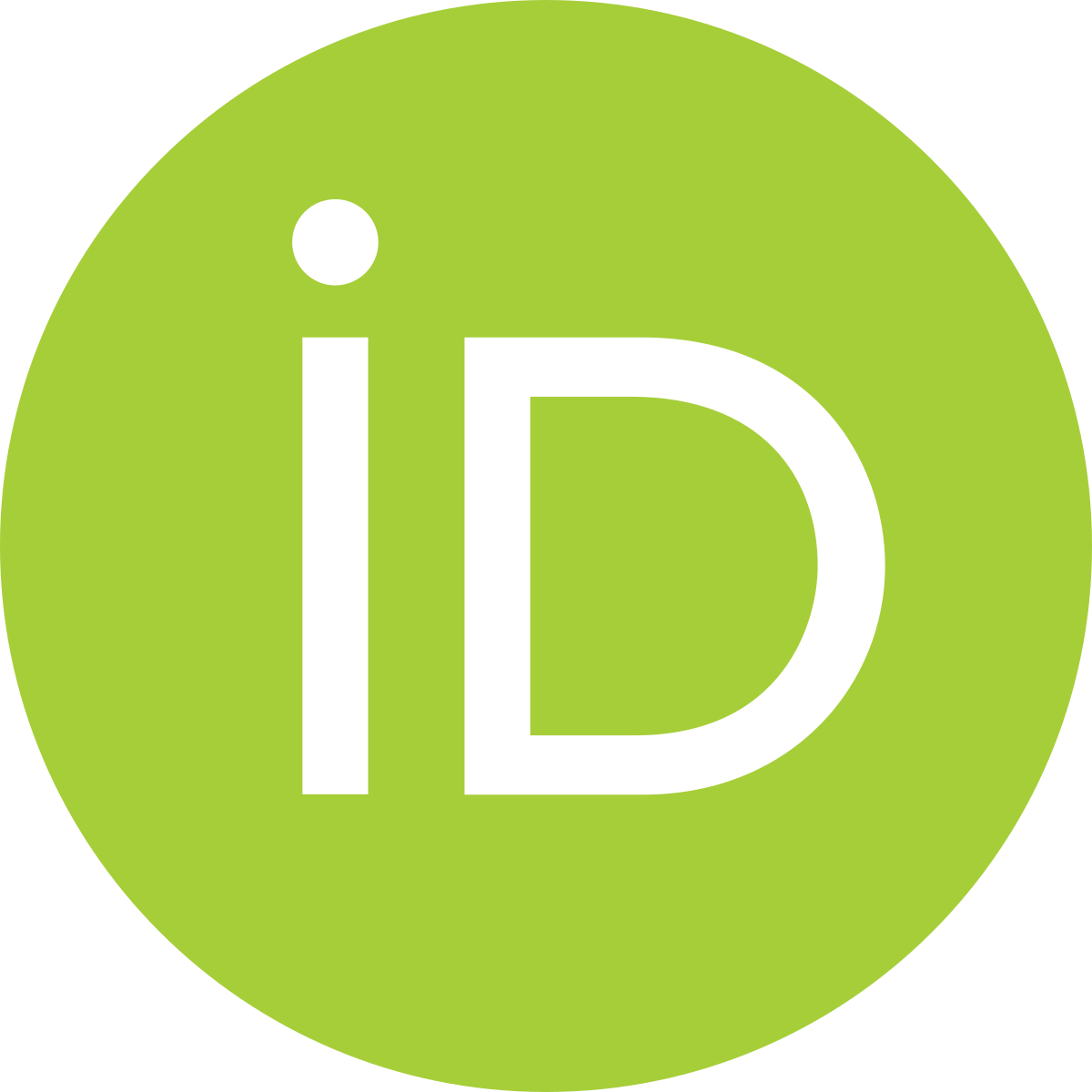}}}

\title{Heavy quarkonia in the light-front quark model}

\author[a]{Muhammad Ridwan\orcid{0000-0002-2949-5866}}
\author*[b,c]{Ahmad Jafar Arifi\orcid{0000-0002-9530-8993}}
\author[a]{Terry Mart\orcid{0000-0003-4628-2245}}

\affiliation[a]{Departemen Fisika, FMIPA, Universitas Indonesia,
   Depok 16424, Indonesia}

\affiliation[b]{Advanced Science Research Center, Japan Atomic Energy Agency (JAEA),
   Ibaraki 319-1195, Japan}
\affiliation[c]{Research Center for Nuclear Physics (RCNP), The University of Osaka, Ibaraki 567-0047, Japan}

\emailAdd{muhammad.ridwan75@sci.ui.ac.id}
\emailAdd{aj.arifi01@gmail.com}
\emailAdd{terry.mart@sci.ui.ac.id}

\abstract{
We present recent progress in constructing heavy quarkonia light-front wave functions from the light-front quark model, a relativistic framework well-suited for describing partonic structure and hadronic form factors. In this proceeding, we discuss the recent application of the model to the $M1$ radiative transition such as $J/\psi \to \eta_c \gamma$ among other transitions. 
Additionally, we compare our results on the transition form factor with available lattice QCD data. 
These show our efforts to deepen our understanding of the structure of heavy quarkonia from light-front model perspective.
}

\FullConference{The 21st International Conference on Hadron Spectroscopy and Structure (HADRON2025)\\
 27 - 31 March, 2025\\
Osaka University, Japan\\}

\begin{document}
\maketitle

\section{Introduction}

Heavy quarkonia, such as charmonium and bottomonium, play a crucial role in exploring the dynamics of Quantum Chromodynamics (QCD) across both perturbative and nonperturbative regimes. 
Traditionally, studies of these mesons utilize wave functions defined in their rest frames; however, high-energy processes, such as deep inelastic scattering, require a description based on light-front dynamics. 
The light-front wave function (LFWF) offers a relativistically consistent framework that connects directly to experimentally accessible observables, especially in high-energy processes.
Despite its advantages, obtaining reliable LFWFs remains a challenge due to the complexity of solving QCD on the light front and the limited availability of model-independent inputs.
While LFWFs can be obtained from light-front Hamiltonian or BLFQ (Basis light-front quantization) approaches~\cite{Li:2017mlw,Li:2018uif}, an alternative method involves solving the rest-frame Hamiltonian and boosting the resulting wave functions to the light-front frame using Melosh rotations to account for relativistic spin effects~\cite{Arifi:2022pal,Arifi:2024mff}.
LFWFs can also be modeled using simple functional forms~\cite{Hwang:2010hw,Li:2021cwv} to facilitate calculations of hadronic observables. 
Moreover, other methods such as light-front holography~\cite{deTeramond:2008ht} and Dyson–Schwinger equations~\cite{Shi:2018zqd} provide useful ways to study hadron structure.

Following this light-front quark model (LFQM) approach~\cite{Arifi:2022pal,Arifi:2024mff}, we constructed the LFWFs of heavy quarkonia and the $B_c$ mesons~\cite{Ridwan:2024ngc}, constrained by their mass spectra and decay constants through statistical analysis.
These LFWFs have been validated through good agreement with available lattice and experimental data for various electromagnetic (EM) properties, including $M1$ radiative transitions such as $J/\psi \to \eta_c \gamma$, among other transitions.
Owing to their distinctive features, such as boost invariance and a simple analytic structure, the resulting LFWFs are useful not only for computing various hadronic form factors but also for describing the partonic structure. 

On the light front, observables are typically calculated using the plus ($\mu=+$) component of the current, where the computation is simpler compared to other components. However, using different components, such as $\mu = \perp$ or $\mu = -$, and different polarization vectors may lead to inconsistent results.
In particular, the minus component is notoriously complicated and is often referred to as the "bad" component.
This difficulty arises because, for instance, $P^-$ is dynamical and includes interaction terms. 
As a result, the minus component of the current depends on the full dynamics of the system, often requiring knowledge of higher Fock states and zero modes.
Nonetheless, the zero-mode issue remains an unresolved puzzle in the light-front framework.  
One of the promising solutions is the replacement of the physical meson mass $M$ by the invariant mass $M_0$ in the matrix element integrand~\cite{Choi:2013mda}. This prescription has been tested in various observables~\cite{Arifi:2022qnd}, including the $M1$ radiative decays in recent studies using different polarization vectors~\cite{Ridwan:2024ngc}.  
This replacement is consistent with the Bakamjian--Thomas construction~\cite{Bakamjian:1953kh}, where the meson state is treated as a noninteracting quark--antiquark system, with interactions incorporated into the mass operator. This approach preserves the Poincaré group structure and commutation relations.

In this proceeding, we present some results for the form factors of $M1$ radiative transitions between heavy quarkonia based on our recent work~\cite{Ridwan:2024ngc} and compare them with lattice QCD calculation~\cite{Dudek:2006ej}. 
We further comment on the operators for the radiative transitions derived from the $+$ and $\perp$ components of EM current in the equal mass constituent quark mass case.
These results reflect our ongoing efforts to provide a more comprehensive understanding of heavy quark systems within the light-front framework.

\section{Light-front quark model}

The meson system at rest is modeled as a bound state of an effectively dressed valence quark and antiquark, satisfying the eigenvalue equation of a QCD-motivated effective Hamiltonian, $H \ket{\psi} = M \ket{\psi}$, where $M$ and $\psi$ are the mass eigenvalue and eigenfunction of the quark--antiquark meson state, respectively.  
We take the Hamiltonian as $H = H_0 + V_{q\bar{q}}$ in the center-of-mass frame of the $q\bar{q}$ system, where $H_0$ is the relativistic kinetic energy operator and $V_{q\bar{q}}$ is an effective interquark potential.  
The potential $V_{q\bar{q}}$ includes confinement, color Coulomb, and hyperfine interactions.  
The eigenfunction $\psi(\bm{k})$ will serve as input for constructing the LFWFs.

The LFWFs can be expressed as
\begin{equation}
\Psi^{JJ_z}_{\lambda_q,\lambda_{\bar{q}}}(x,\bm{k}_{\perp}) = \Phi(x,\bm{k}_{\perp}) \, \mathcal{R}^{JJ_z}_{\lambda_q,\lambda_{\bar{q}}}(x,\bm{k}_{\perp}),
\end{equation}
where $\Phi(x,\bm{k}_{\perp})$ is the radial wave function and $\mathcal{R}^{JJ_z}_{\lambda_q,\lambda_{\bar{q}}}(x,\bm{k}_{\perp})$ is the spin-orbit wave function obtained through the Melosh transformation.  
The spin-orbit wave functions $\mathcal{R}^{JJ_z}_{\lambda_q \lambda_{\bar{q}}}$ satisfy the unitary condition.
The LFWF is defined using Lorentz-invariant internal variables: $x_i = p^+_i / P^+$, $\bm{k}_{\perp i} = \bm{p}_{\perp i} - x_i \bm{P}_{\perp}$, and light-front helicity $\lambda_i$.  
Here, $P^\mu = (P^+, P^-, \bm{P}_\perp)$ is the meson four-momentum, and $p^\mu_i$ is the four-momentum of the $i$-th constituent quark ($i=1,2$), with $x \equiv x_1$ and $\bm{k}_\perp \equiv \bm{k}_{\perp 1}$.  
The internal three-momentum is $\bm{k} = (k_z, \bm{k}_\perp)$, related to light-front variables by $k_z = \left(x - {1}/{2}\right) M_0$, where the boost-invariant squared mass is $M_0^2 = ({\bm{k}_\perp^2 + m_q^2})/{x(1-x)}$.  
The variable change $\{k_z, \bm{k}_\perp\} \to \{x, \bm{k}_\perp\}$ introduces the Jacobian ${\partial k_z}/{\partial x} = {M_0}/{4x(1-x)}$.

The radial wave function can be written as $\Phi(x,\bm{k}_\perp) = \sqrt{2(2\pi)^3} \sqrt{\partial k_z/\partial x} \, \psi(\bm{k})$, where $\psi(\bm{k})$ can be Gaussian, harmonic oscillator (HO), power-law type, or an accurate eigenfunction of the rest-frame Hamiltonian.  
Another common form of $\Phi(x,\bm{k}_\perp)$ is based on the Brodsky-Huang-Lepage (BHL) prescription.  
We use a few HO basis functions important for describing excited states (see Ref.~\cite{Ridwan:2024ngc} for details).  
The Jacobian factor $\partial k_z / \partial x$ ensures proper normalization, so that the LFWFs satisfy
$\int [\dd^3 \tilde{k}]\, |\Phi(x,\bm{k}_\perp)|^2 = 1,$
where the shorthand notation is defined as $[\dd^3 \tilde{k}] \equiv {\dd x\, \dd^2 \bm{k}_\perp}/{2(2\pi)^3}$.

\section{Application: $M1$ radiative transitions} 
\label{sec:radiative}

As one of the applications of our LFWFs~\cite{Ridwan:2024ngc}, here we analyze the $M1$ radiative transitions between vector and pseudoscalar heavy quarkonia where the transition form factor  $F_{\mathcal{VP}}(q^2)$ can be defined as~\cite{Jaus:1991cy,Choi:2007se}
\begin{eqnarray}
\label{eq:rad_M1}
\bra{\mathcal{P}(P^\prime)} J_{\rm em}^\mu(0)\ket{\mathcal{V}(P,h)} = ie \varepsilon^{\mu \nu \rho\sigma} \epsilon_\nu q_\rho P_\sigma F_\mathcal{VP}(q^2),\quad 
\end{eqnarray}
where the antisymmetric tensor $ \varepsilon^{\mu \nu \rho\sigma} $ assures EM gauge invariance, and $q = P-P^\prime$ is the four-momentum of the virtual photon. 
The matrix element $\mathcal{J}_h^\mu\equiv\bra{\mathcal{P}(P^\prime)} J_{\text{em}}^{\mu} \ket{\mathcal{V}(P,h)}$ is computed in the $q^+=0$ frame and can be expressed in terms of the convolution formula of the initial and final LFWFs as~\cite{Jaus:1991cy,Choi:2007se}
\begin{eqnarray}
\mathcal{J}_h^\mu = \sum_j e e_q^j \int [\dd^3 \tilde{k}]~\Phi(x, \bm{k}_\perp^\prime) \Phi(x,\bm{k}_\perp) \sum_{\lambda,\lambda^\prime,\bar{\lambda}} \mathcal{R}_{\lambda^\prime \bar{\lambda}}^{00\dagger}(x, \bm{k}_\perp^\prime) \frac{\bar{u}_{\lambda^\prime}(p_1^\prime) }{\sqrt{x}} \gamma^{\mu} 
\frac{u_{\lambda}(p_1) }{\sqrt{x}} \mathcal{R}_{\lambda \bar{\lambda}}^{1h}(x, \bm{k}_\perp),
\end{eqnarray}
where $e_q^j$ is the electric charge for $j$-th quark flavor.
By matching the left- and right-hand sides of Eq.~\eqref{eq:rad_M1},
the form factor can be extracted from $F_\mathcal{VP}(q^2) = {\mathcal{J}_h^\mu}/{\mathcal{G}_h^\mu}$
with $\mathcal{G}_h^\mu=ie \varepsilon^{\mu \nu \rho\sigma} \epsilon_\nu q_\rho P_\sigma$, where the mass $M$ should be replaced by $M_0$ and inserted in the integrand.
We note that the form factor receives the contribution from two processes where the photon couples either to the quark or antiquark.
Thus, the from factor can be computed as
\begin{eqnarray}
\label{eq:operator}
 F^{\mu}_{h}(Q^2) = 2e_q \int [\dd^3 \tilde{k}] \frac{\Phi(x,\bm{k}_\perp^\prime) \Phi(x,\bm{k}_\perp)}{\sqrt{m_q^2+\bm{k}_\perp^{\prime 2}} \sqrt{m_q^2+\bm{k}_\perp^2}  } \mathcal{O}_{\mathcal{VP}\gamma}^\mu(h),
\end{eqnarray}
where the operators can be computed by using various combination of current components and polarization vectors. 
For two different polarizations, the operators are given by~\cite{Ridwan:2024ngc}
\begin{align}
     \mathcal{O}_{\mathcal{VP}\gamma}^+(\pm1) = 2(1-x)\left(m_q + \frac{\bm{k}^2_{\perp}}{\mathcal{D}_0}\right), \qquad
 \mathcal{O}_{\mathcal{VP}\gamma}^{R(L)}(0) = \frac{m_q}{xM_0}\left(m_q +\frac{2\bm{k}^2_{\perp}}{\mathcal{D}_0}\right),  
\end{align}
where the longitudinal polarization can only be extracted using $J^\mu=J^{R(L)}=J^x\pm J^{iy}$.
The coupling constant related to the decay process is computed as $g_{\mathcal{VP}\gamma}=F_{\mathcal{VP}\gamma}(q^2=0)$, from which we can compute the decay width. 


\begin{figure}[b]
	\centering
	\includegraphics[width=1\columnwidth]{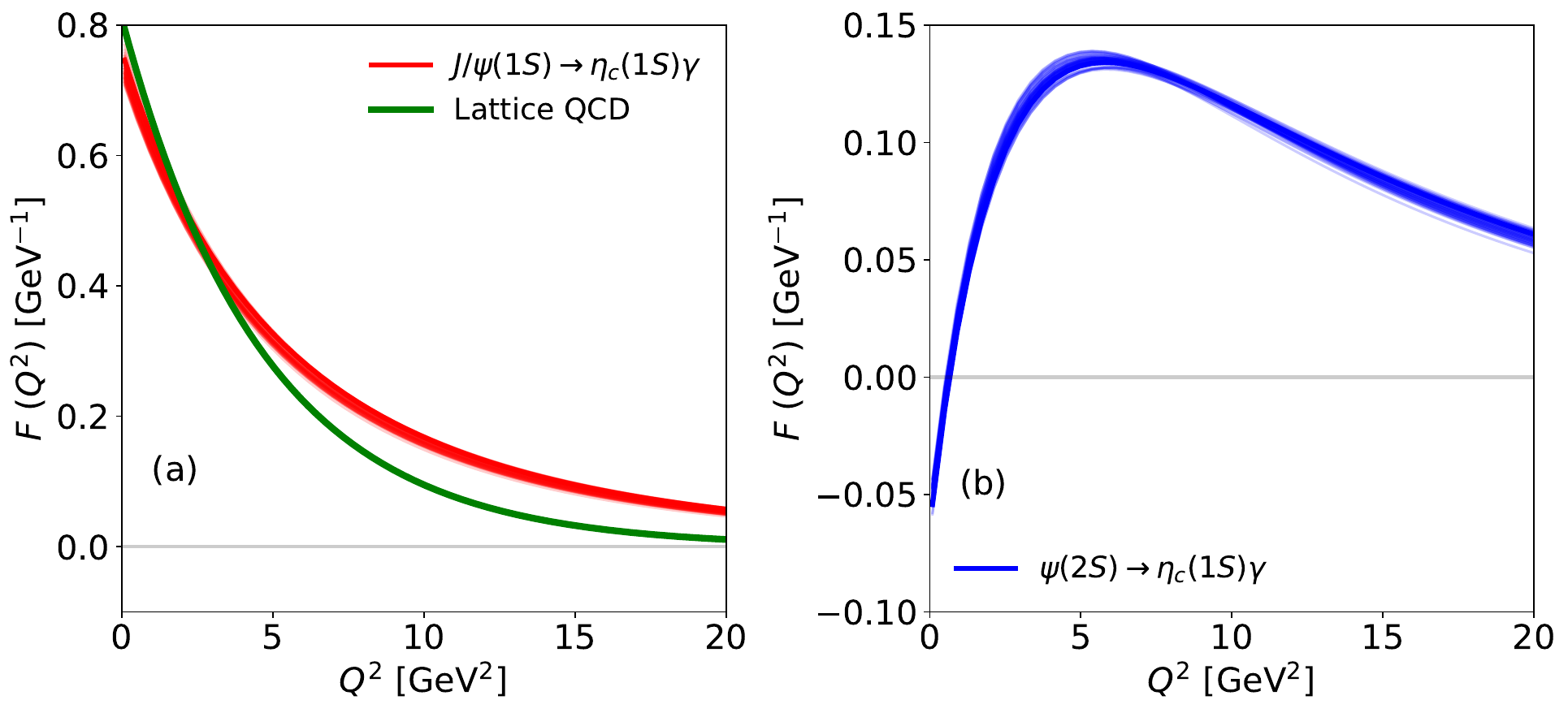}
	\caption{\label{fig:form_factor} Transition form factors of heavy quarkonia for allowed and hindered cases, shown in the left and right panels, respectively. The $J/\psi\to\eta_c\gamma$ transition form factor is compared with a fitted function to reproduce the lattice QCD data~\cite{Dudek:2006ej}. We note that the lattice QCD data have been converted to our notation in Eq.~\eqref{eq:rad_M1}.}
\end{figure}

Although the operators in Eq.~\eqref{eq:operator} are different, the resulting form factor are exactly the same.
We note that the $M$ replaced by $M_0$ in the second operator $\mathcal{O}_{\mathcal{VP}\gamma}^{R(L)}(0)$ is essential to obtain the consistency. 
The HO basis function is inherently rotational invariant which help obtaining the consistency for different polarization vectors.
We note that the anisotropic LFWF, characterized by $\langle 2k_z^2 \rangle \neq \langle k_\perp^2 \rangle$, can lead to different results depending on the polarization vector.
Furthermore, we expect also the operator with minus component will give the same result, similar to other observables although some modification of Lorentz structure may be required.
The numerical confirmation is left for the future studies.

Compared to the conventional quark model~\cite{Godfrey:1985xj}, one of the main advantages of using the LFWFs is its application for computing various form factors up to high-energy momentum transfer.
The transition form factors $F_{\mathcal{VP}\gamma}(Q^2)$ for representative processes $\mathcal{V}(nS)\to \mathcal{P}(n^\prime S)+\gamma$ are presented in Fig.~\ref{fig:form_factor}, where the allowed ($n=n^\prime$) and hindered ($n\neq n^\prime$) processes are given in the left and right panels, respectively.  
As $Q^2\to 0$, $F_{\mathcal{VP}\gamma}(Q^2)$ are significantly enhanced for the allowed case since they are established from an overlap of the wave functions of the same principal and partial wave $nS$.  
Meanwhile, they are much suppressed for the hindered case due to the orthogonality of the initial and final wave functions.
In the left panels of Fig.~\ref{fig:form_factor} for the allowed case, we find good agreement between our result on $F_{J/\psi\eta_c\gamma}(Q^2)$ with a fitted function to the lattice QCD data~\cite{Dudek:2006ej}, although the deviation is seen in the higher $Q^2$ region.
In Table~\ref{tab:coupling}, we also compare our results of the coupling constants for selected process with available experimental data~\cite{ParticleDataGroup:2024cfk} and other models such as BLFQ~\cite{Li:2018uif} and Godfrey-Isgur (GI)~\cite{Godfrey:1985xj}. 
Our results show reasonable agreement with these data and are useful for guiding future lattice QCD simulations or experiments.

\begin{table}[t]
	\centering
	\caption{Coupling constants calculated in the present work for selected transitions compared with experimental data and those obtained from other models, such as BLFQ~\cite{Li:2018uif} and GI~\cite{Godfrey:1985xj}.}
	\label{tab:coupling}
	\begin{tabular}{l|cccc}
		\hline\hline
		$g_{\mathcal{VP}\gamma}$ [GeV$^{-1}$] & $J/\psi \rightarrow \eta_c\gamma$ & $\psi^\prime \rightarrow \eta_c\gamma$ & $\Upsilon \rightarrow \eta_b \gamma$ & $\Upsilon^\prime \rightarrow \eta_b \gamma$ \\
		\hline
		Our & 0.745(15) & $-0.0605(37)$ & $-0.1279(9)$ & 0.0049(1) \\
		Exp.~\cite{ParticleDataGroup:2024cfk} & 0.684(85) & $-0.040(3)$ &\dots & 0.0057(6) \\
        BLFQ~\cite{Li:2018uif} & 0.873 & $-0.144$ & $-0.141$ & 0.011\\
        GI~\cite{Godfrey:1985xj} &  0.690 & $-0.056$ & $-0.130$ & 0.007 \\
		\hline \hline
	\end{tabular}
\end{table}

\section{Conclusion and Outlook}

We have presented our progress in studying various form factors of heavy quarkonia using the light-front quark model.
The light-front wave functions (LFWFs) were constructed based on a variational approach combined with the Melosh rotation.
Model parameters were fitted to experimental mass spectra and decay constants, yielding results in reasonable agreement with the data.
We have applied our LFWFs to the $M1$ radiative transitions and found consistency between the results obtained using different polarization vectors and the available experimental and lattice data.
Notably, the replacement $M \to M_0$ was found to be important for achieving such self-consistent results.
For future work, we plan to apply our LFWFs to other observables, refine the model wave functions for improved predictive power, and analyze different current components to gain deeper insights into the structure of heavy quarkonia within the light-front framework.

\section*{Acknowledgments}

A.J.A. is supported by JAEA Postdoctoral Fellowship Program and was partly supported by the RCNP Collaboration Research Network program under project number COREnet 057. 
T.M. is supported by the PUTI Q1 Grant from University of Indonesia under contract No. NKB-441/UN2.RST/HKP.05.00/2024.

\end{document}